\documentclass[sigconf]{acmart}

\renewcommand\footnotetextcopyrightpermission[1]{}  
\acmConference[Preprint]{ArXiv Preprint}{September 2024}{}

\usepackage{booktabs}
\usepackage{listings}
\usepackage{quoting}
\usepackage{listings}
\newcommand{\qq}[1]{\begin{quote}\textit{#1}\end{quote}}
\quotingsetup{leftmargin=1.5em, rightmargin=1.5em, vskip=0pt}

\begin{document}

\title{ELLMA-T: an Embodied LLM-agent for Supporting English Language Learning in Social VR }

\author{Mengxu Pan}
\email{pan.meng@northeastern.edu}
\affiliation{%
  \institution{Khoury College of Computer Science, Northeastern University}
  \city{Vancouver}
  \country{Canada}
}

\author{Alexandra Kitson}
\email{akitson@sfu.ca}
\orcid{0003-3479-5297}
\affiliation{%
  \institution{Simon Fraser University}
  \city{Vancouver}
  \state{BC}
  \country{Canada}
}

\author{Hongyu Wan}
\email{wan.hongy@northeastern.edu}
\affiliation{%
  \institution{Khoury College of Computer Science, Northeastern University}
  \city{Vancouver}
  \country{Canada} 
}

\author{Mirjana Prpa}
\email{m.prpa@northeastern.edu}
\affiliation{%
  \institution{Khoury College of Computer Science, Northeastern University}
  \city{Vancouver}
  \country{Canada}
}

\begin{abstract}
Many people struggle with learning a new language, with traditional tools falling short in providing contextualized learning tailored to each learner's needs. The recent development of large language models (LLMs) and embodied conversational agents (ECAs) in social virtual reality (VR) provide new opportunities to practice language learning in a contextualized and naturalistic way that takes into account the learner's language level and needs. To explore this opportunity, we developed ELLMA-T, an ECA that leverages an LLM (GPT-4) and situated learning framework for supporting learning English language in social VR (VRChat). Drawing on qualitative interviews (N=12), we reveal the potential of ELLMA-T to generate realistic, believable and context-specific role plays for agent-learner interaction in VR, and LLM's capability to provide initial language assessment and continuous feedback to learners. We provide five design implications for the future development of LLM-based language agents in social VR.
\end{abstract}

\keywords{Embodied Conversational Agents, Social Virtual Reality, Large Language Models, Language Learning}

\begin{teaserfigure}
  \includegraphics[width=\textwidth]{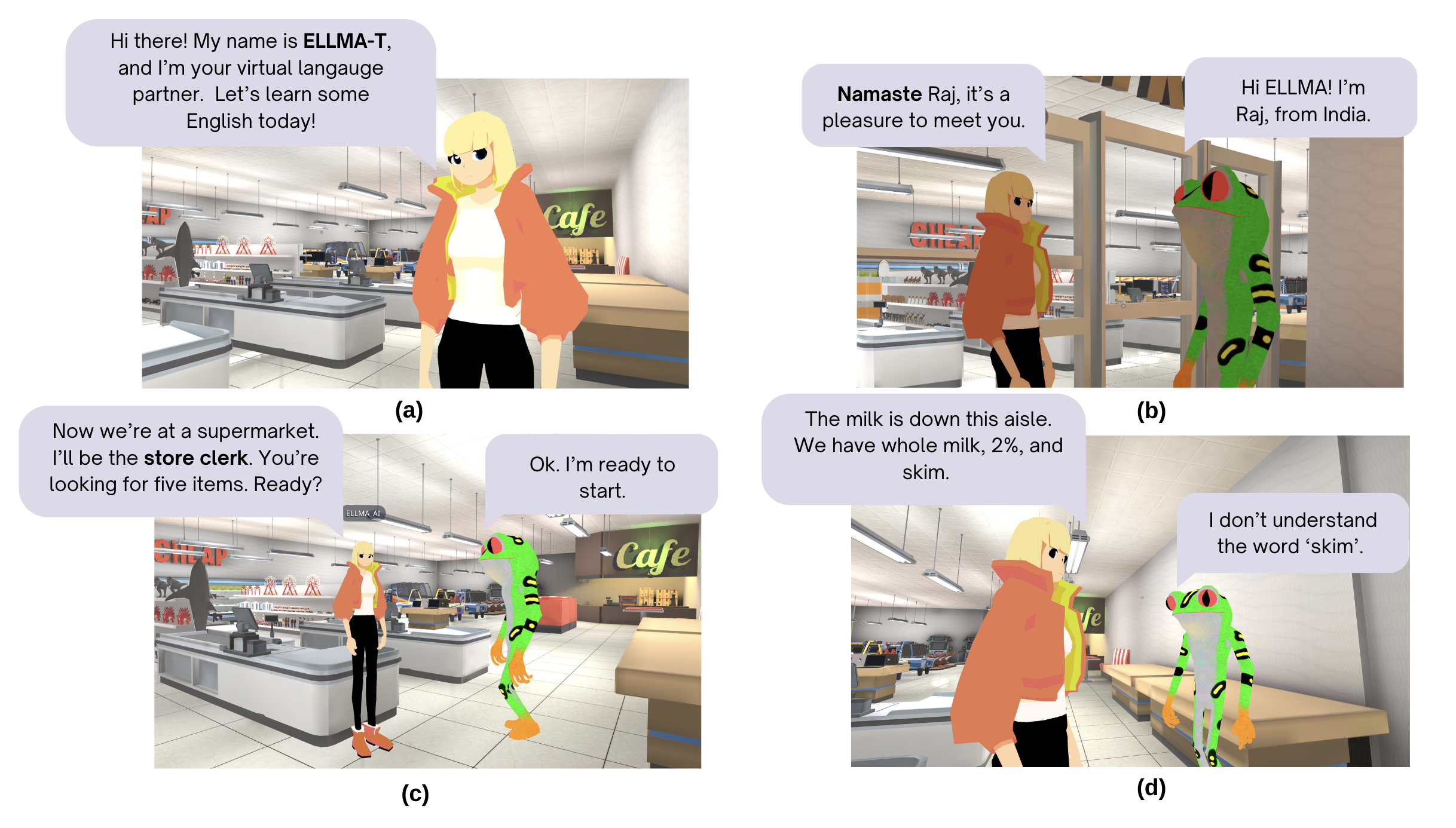}
  \caption{a) ELLMA-T, the GPT-4 based language tutor in VRChat; b) b) ELLMA-T greeting the user in their native language; c) Role-play scenario with ELLMA-T and the user in a supermarket setting; d) User asking ELLMA-T for word explanations.}
  \Description{A teaser figure shows the introduction of ELLMA-T, a GPT-4 based ECA in VRchat in a role-play at a supermarket scenario.}
  \label{fig:teaser}
\end{teaserfigure}

\maketitle

\section{Introduction}
Acquiring proficiency in English as an additional language is essential for many facets of life. This applies to both countries where English is taught as part of the compulsory curriculum, even though it is not the native language, and to English-speaking countries where different people, such as immigrants, visitors, tourists or international students require additional support to practice and improve their English skills \cite{adjagbodjou_envisioning_2024}. Commonly used learning approaches administered in traditional classrooms often fall short of providing a contextualized learning environment, tailored learning materials, and engaging experiences to support learning.

Besides commonly used language learning tools (e.g., mobile applications such as Duolingo \cite{duolingo_introducing_2023}), the recent proliferation of social Virtual Reality (VR) platforms presents new learning channels. VRChat \cite{vrchat_vrchat_2024} has become a popular social platform for online interactions in a virtual environment with over 48k daily active users and \~8.8 million total. Though not specifically designed for foreign language learning, players take advantage of its immersive environment to have authentic conversation experiences with native speakers \cite{merle_meruru555_en-jp_2024}. Previous research has shown social interactions in VR to be positive and helpful for learning a foreign language \cite{peterson_efl_2012, cho_research_2022}. Meanwhile, players encounter numerous obstacles when undertaking such endeavors. For example, it can be difficult and time-consuming to find a native speaker to interact with; and even when found, native speakers may not be willing or suitable to act as language tutors. Moreover, users may experience increased social anxiety during the search and interaction and overall the learning experience is deemed to be unsystematic \cite{roy_and_aimee_can_2020, jouzu_juls_vrchat_2022}.

The recent emergence of LLM-based embodied conversational agents (ECA) in social VR, such as Celeste AI \cite{ko-ficom_celeste-ai_2024}, has received significant attention from VRChat users because they converse and behave in human-like ways. This presents an immediate opportunity to leverage LLM-agents for language learning in social VR. While LLMs have been explored for various educational purposes \cite{kasneci_chatgpt_2023}, their application in language learning, especially in social VR contexts, remains underexplored. To that end, we present our efforts to answer the following research question (RQ):\textit{ How can we leverage situated learning and an LLM-based ECA for language learning in social VR?}

We built ELLMA-T, an embodied LLM-agent to explore the potential of using LLMs and situated learning framework \cite{brown_situated_1989, lave_situated_1991} to facilitate language learning in social VR and conducted a user study (N=12) with ELLMA-T using semi-structured interviews. By leveraging an LLM-based agent in VRChat, the system aims to provide an authentic conversation experience and personalized feedback to users, enhancing the contextualized language learning experience. The contributions we present in this paper are the following: 
\begin{enumerate}
    \item Design of ELLMA-T, an embodied LLM-agent tasked to help adult language learners from A1 to C1 level practice spoken English, by simulating an immersive real-life conversation within social VR;
    \item A user study and findings about participants' perception of ELLMA-T's capabilities and limitations on four English learning tasks: language level assessment, role-play dialogue generation, feedback generation, and scaffolding abilities, contributing to the larger picture of LLM-assisted language learning and LLM's applications;
    \item Design implications for designing embodied LLM-agents in social VR context for language learning.
\end{enumerate}

Our findings from a preliminary user study demonstrate ELLMA-T's potential as a 'human-like' language tutor for situated language learning. However, challenges such as conversation flow disruptions and insufficient emotional support were identified, suggesting areas for future improvement. The study’s design implications extend to the design of personalized, culturally relevant, and adaptive learning systems that integrate verbal and non-verbal cues. These findings highlight the potential for further exploration into developing LLM-based ECAs in social VR for long-term language learning. 

\section{Related Work}
\subsection{Supporting Situated Language Learning in VR }

Situated learning theory is grounded in the understanding that the most effective learning happens in context. Learning can neither be separated from the specific situations in which it occurs nor be transferred or applied fully in a school setting \cite{brown_situated_1989, lave_situated_1991}. In language acquisition, an authentic context provides learning opportunities that closely simulate real-life situations and includes cultural nuances crucial for meaningful and engaging language use \cite{lan_chapter_2020}, for example, discerning the nuances and subtleties of word usage in polysemous situations \cite{wong_cognitive_2018}. 

The immersive nature of VR is particularly suitable for creating a realistic language learning environment and simulating interpersonal interactions in context-rich scenarios. Since 2013 and the expansion of affordable head-mounted displays \cite{dhimolea_systematic_2022,limniou_full_2008}, VR has been widely utilized in many areas of training and education \cite{aim_effectiveness_2016,zhu_learniotvr_2023}. Within VR, 3D objects and avatars have been used to facilitate embodied and contextualized learning experiences \cite{chatain_three_2023}. With rapid advancements in VR headsets, learners experience a heightened sense of presence, increased agency, and embodiment that support learning comparable to learning-by-doing in the physical world \cite{kaplan-rakowski_one--one_2021}.

A growing body of literature has investigated the usage of VR for English as foreign language (EFL) learning, specifically how VR environments can be used for different language learning tasks \cite{lin_language_2015,klimova_use_2021,dhimolea_systematic_2022}. Research indicates that VR is beneficial for vocabulary acquisition, communication skills, and writing skills \cite{lan_chapter_2020}. Other benefits of using VR for EFL learning include increased learners' motivation, engagement, enjoyment, automation \cite{makransky_structural_2018,lan_chapter_2020, dhimolea_systematic_2022}, and reduced anxiety \cite{gruber_user_2020}. EFL learning using VR can also benefit kids with complex needs and low-achieving students \cite{lan_chapter_2020}.

Researchers have generated several strategies to enhance learning in VR by contextualization and extending memory retention. These strategies include interacting with a 3D object for vocabulary understanding \cite{fuhrman_moving_2021,hsu_spelland_2023}, completing tasks in real-life scenarios \cite{wik_embodied_2009}, problem-based learning \cite{chen_virtual_2021}, and integrating learning into narrative or through gamification \cite{cheng_teaching_2017,hsu_spelland_2023,seow_lingoland_2023}. In addition, the use of avatars in VR has been shown to support human-to-human collaboration, and enhance a sense of community during the learning experience \cite{jackson_collaboration_2000,peterson_efl_2012,cantone_contextualized_2023}. The use of avatars may in particular support often overlooked nuances of learning languages such as the role of gestures that accompany spoken language. For example, \cite{cheng_teaching_2017} showed how language usage and body movement are integrated to teach learners intercultural nuances, such as the practice of greeting in Japanese while simultaneously bowing, demonstrating how verbal and non-verbal cues work together in communication.

\subsection{Embodied Conversational Agents for Language Learning}

Many VR language learning applications involve embodied conversational agents (ECAs) in the environment. An ECA is a digitally animated character capable of engaging in natural, human-like communication with users \cite{cassell_embodied_2000}. As Cassell states, an ideal ECA is described as being able to conduct a human-to-agent conversation in a way that mimics human face-to-face communication so that it feels \textit{believable}. Many other features have been discussed for the design of the ECAs such as the use of verbal and non-verbal cues, multimodal communication, appearance and visualization \cite{cassell_embodied_2000}, and carefully crafted personality traits \cite{van_mulken_persona_1998}, all aiming to increase agent credibility. In recent studies, ECAs are increasingly considered as social agents \cite{atkinson_fostering_2005}, designed not only to interact but also to understand and adapt to the social and cultural contexts of the user. 

Recent studies have noted that ECAs, despite their potential, often fall under the broader category of chatbots, which lack a unified conceptual framework and typically do not include a virtual body or multimodal processing abilities \cite{jeon_beyond_2023}. Meanwhile, much research has focused on the design and use of conventional chatbots (no embodiment) for language learning, but the exploration of ECAs remains limited. Early efforts, such as those by \citet{wik_embodied_2009}, employed humanoid agents to perform different language learning tasks, demonstrating success across diverse cultural backgrounds. \citet{divekar_foreign_2022} also used humanoid agents to mimic real-life scenarios, specifically non-dyadic (multiple-people) multi-modal conversations. More recent studies have shown that learners can develop strong emotional bonds with ECAs, leading to increased motivation and positive learning outcomes \cite{jeon_exploring_2022}. \cite{xu_same_2021} demonstrating also a positive correlation between learners' strong affection for agents' appearances and improved learning outcomes.

Despite the potential of ECAs for simulating human-like behaviour, much of the existing research on ECAs for language learning features ECAs with minimal interactive abilities \cite{wik_embodied_2009, divekar_foreign_2022, jeon_exploring_2022}. These ECAs are often restricted to basic functions like lip-syncing, and limited non-verbal gestures and expressive behaviour due to technological constraints. The recent advancements in large language models (LLMs) provide new technological capabilities for ECAs, opening up design opportunities for more dynamic, interactive ECAs.

\subsection{Potential of LLMs \& LLM-agents for Language Learning in Social VR}

Previous natural language processing (NLP) techniques had been applied to English language learning tools, which usually require templates or use pre-scripted goal-oriented scenes \cite{tyen_towards_2022}. Previous (E)CAs' responses regarding language learning and feedback also remained limited \cite{zhang_justspeak_2021}. This limitation was lifted with the recent development of LLMs and their content generation ability that extends beyond language tasks \cite{yang_harnessing_2024}. In addition to language generation tasks, LLMs can be leveraged for generating ECAs' 3D avatars \cite{hong_avatarclip_2022}, controlling motion \cite{qing_story--motion_2023}, generating motion and facial expressions \cite{normoyle_using_2024}, and driving ECA's behavior \cite{normoyle_using_2024, kumaran_scenecraft_2023}. Multimodal LLMs have the potential to understand the virtual environment and the objects in it \cite{wang_chat-3d_2023}, for even more contextualized learning.

While the number of discussions around the usage of LLMs is growing, particularly chatGPT for education across disciplines \cite{kasneci_chatgpt_2023}, fewer studies focus on the use of LLMs for language learning. Previous studies have explored the use of LLM-based chatbots in academic settings for EFL for tasks like grammar correction and essay writing \cite{zhang_chatbot-based_2023, law_application_2024}. In a recent study, \cite{caines_application_2023} examined the use of LLMs in language teaching and assessment, including the assessment of generated content, calibration of existing teaching materials, automated assessment of language learners, and providing feedback. \cite{caines_application_2023} review of LLMs besides GPT suggests that alternative models for content generation may be a better choice due to their multilingual capabilities and their general availability as open-source models; yet, for other tasks like providing feedback to language learners, the usage of GPT models is still promising.

Despite the advancements of LLMs, the use of LLM-based ECAs in language learning applications remains relatively underexplored. Recent studies have begun to explore how personality expression and embodiment in LLM-based educational agents impact learning outcomes \cite{sonlu_effects_2024}. Yet, there remains a gap in understanding the complexity of factors that influence LLM-based agent-supported language learning in social VR. To address this gap, we explore the design and integration of LLM-based ECAs into social VR for contextualized language learning.

\section{System Design}
Our system implements an LLM-based agent to assist learners in practicing spoken English within a social VR environment. The agent (ELLMA-T) engages with the users in role-play conversations based on real-life scenarios, providing interactive and adaptive language practice in VRChat. In the proceeding section, we discuss our design principles based on education theory, learning task design, and system implementation.

\subsection{Design Principles}

We based the following system design principles on prior research involving additional language learning regardless of the learner's age or academic status:
\begin{enumerate}
    \item \textbf{Student-centered learning}, with the system designed to fit each learner's unique needs \cite{darsih_learner-centered_2018};
    \item The learning material follows \textbf{task-oriented learning}, meaning the learner needs to complete a learning-related task rather than engaging in non-task-based casual conversations \cite{raux_using_2004};
    \item The system focuses on providing \textbf{emotional support} to encourage learning, a major factor in sustaining continuous learning for language learners \cite{dincer_common_2013}.
\end{enumerate}

We build on the needs for supporting adult English language learning put forth by \citet{adjagbodjou_envisioning_2024}: (a) daily conversational needs that are contextualized, (b) tools that address both linguistic and emotional needs, and (c) long-term, contextually situated learning experiences. Though this system is designed with adult learners' needs in mind, our goal is to scale the system for different user groups of various ages. First, we emphasize creating an immersive environment with an embodied agent in VR to simulate real-life conversation experiences, addressing need (a). Second, we explore different communication strategies and channels that address learners' emotional needs using LLMs' language abilities, addressing need (b). Third, we use VRChat as a free and easily accessible open social platform, which provides a rich space for users to connect, form communities, and potentially pursue collaborative learning opportunities, addressing need (c).

\subsection{Learning English with ELLMA-T in VRChat}

\begin{figure*}[ht]
  \centering
  \includegraphics[width=\linewidth]{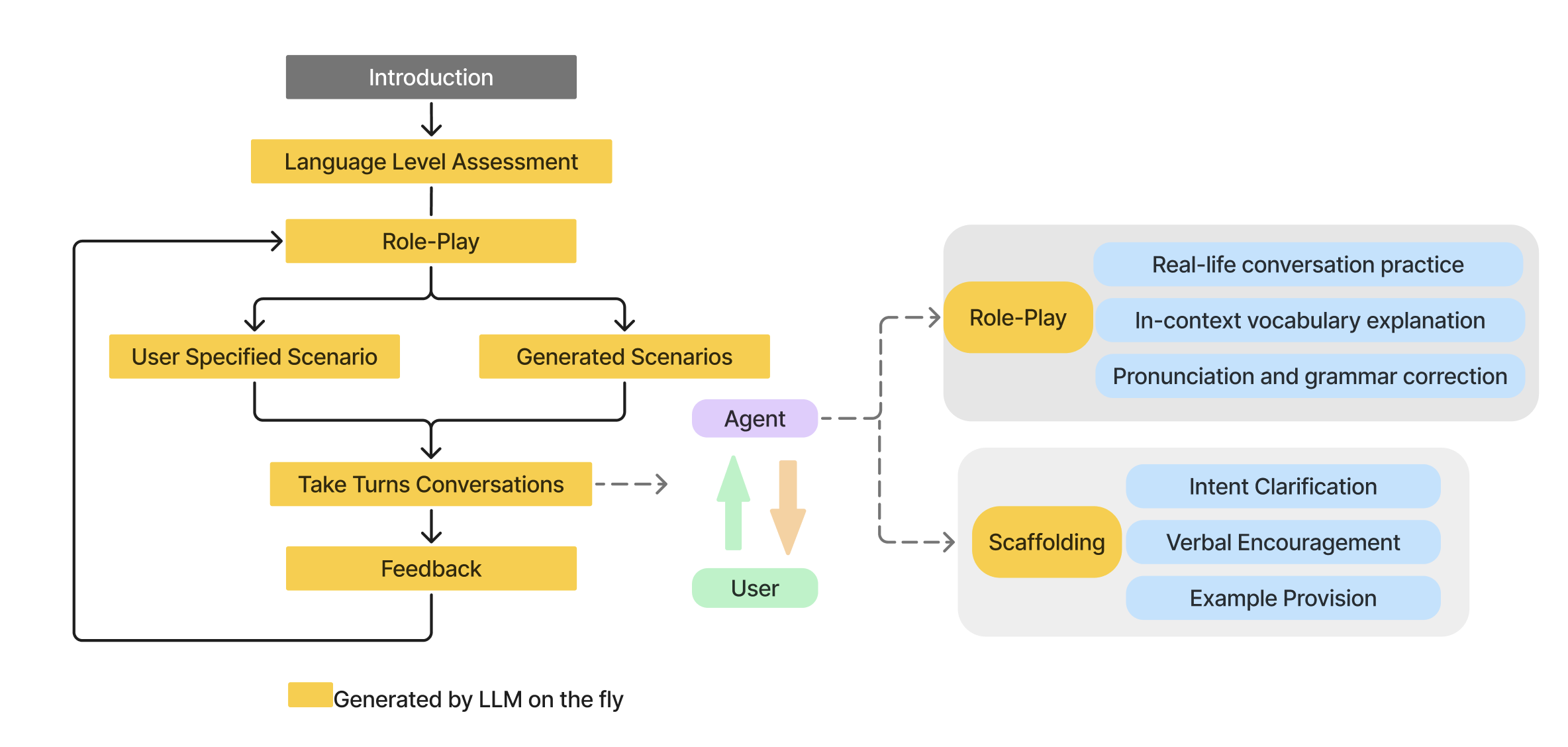} 
  \caption{Workflow of conversation tasks performed by ELLMA-T, including greeting the user, conducting language assessments, engaging in role-play scenarios, and providing feedback.}
  \Description{This flowchart illustrates the workflow and tasks executed by the LLM-based system for language learning.

The flowchart begins with the "Introduction" step. Then, the second step, "Language Level Assessment". Then the system moves to the "Role-Play" stage. Here, the flow branches into two paths: user-specified scenarios or generated scenarios. Both branches lead to the "Take Turns Conversations" step. Following the conversation, the system provides "Feedback" to the user.

The agent and user interaction is shown through a double-headed arrow indicating continuous engagement between the agent and user throughout the process.
Role-Play: This step involves real-life conversation practice, in-context vocabulary explanation, and pronunciation and grammar correction.
Scaffolding: This involves intent clarification, verbal encouragement, and example provision to support the user during the conversation.

Color Coding:
Steps highlighted in yellow indicate that they are generated by the LLM on the fly.
The right side of the flowchart is color-coded to differentiate between the Role-Play and Scaffolding tasks, each with its sub-tasks.
}
  \label{fig:conversation_tasks}
\end{figure*}

The workflow (see Fig. \ref{fig:conversation_tasks}) starts with ELLMA-T asking for the learner's information and assessing their language level. ELLMA-T then generates a conversation scenario specified by the learner or offers three scenarios for the learner to choose from. During role-play conversation practices, ELLMA-T acts as the language practice partner, playing the character in the scene and providing scaffolding when needed. After the role-play, ELLMA-T provides feedback to the learner based on the conversation in the role-play. The learner can end the conversation or switch to another role-play at any point in the conversation. Informed by our three design principles, we designed ELLMA-T's language level assessment, conversation learning materials, and communication strategy, which were generated by the LLM on the fly using prompt engineering (see Section \ref{system_implementation}).

\subsubsection{Design Principle \#1: Introduction and Language Level Assessment}
\textit{Introduction:} ELLMA-T initiates the conversation by introducing itself as a friendly and culturally aware language partner. It gathers information from the user through open-ended questions about their name, cultural background, motivation to learn English, and any additional details they wish to share to form a sense of personal connection. For example, ELLMA-T greets the learner in their native language and acknowledges their culture and personal experiences. During the first few rounds of conversations, the learner gets familiar with the agent and the VR environment.   

\textit{Language Level Assessment Task:} Next, ELLMA-T gives the learner a topic to speak about to assess the learner's current level of English proficiency. The topic and the format of the free speech assessment follow the format of an IELTS \cite{ielts_ielts_2024} speaking test. Whereas IELTS is a more formal conversation, in our system the assessment is a casual conversation. In the following learning session, ELLMA-T adjusts its language and topic difficulty according to the learner's proficiency. The length of the free speech varies from 30 seconds to as long as the learner continues speaking on the topic. The agent makes the judgment if it has long enough input to assess the learner's spoken language. If the user's input is not long enough, the agent asks follow-up questions and encourages the learner to speak more. Once ELLMA-T decides it has enough input from the learner, it assesses the learner's spoken language proficiency and gives a result from A1 to C2 based on CEFR criteria \cite{coe_cefr_2024}. 

\subsubsection{Design Principle \#2: Role-Play Topic Generation and Conversation Continuation}

\begin{figure*}[ht]
  \centering
  \includegraphics[width=\linewidth]{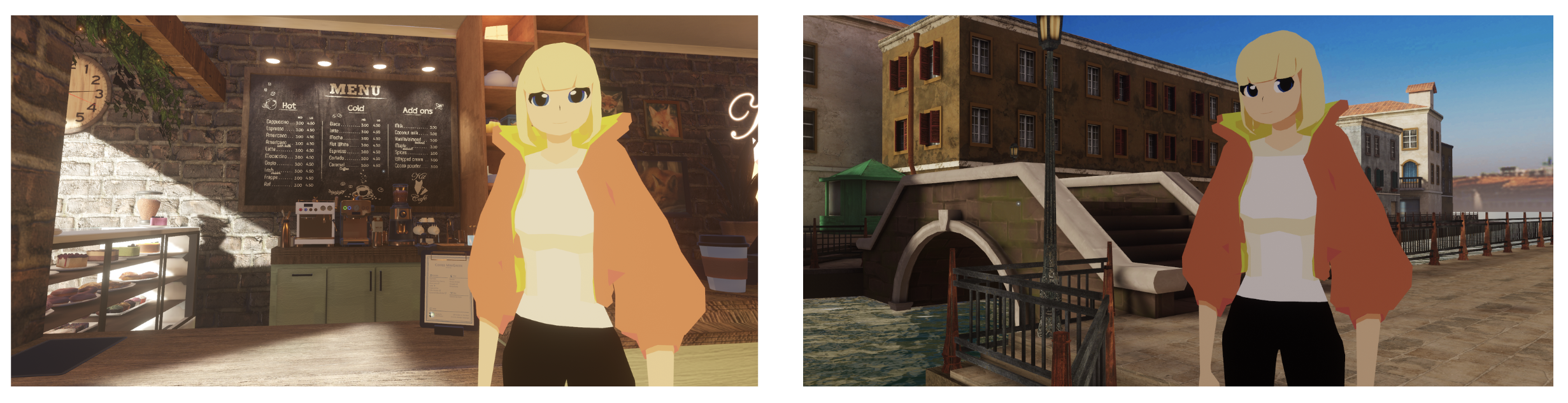} 
  \caption{ELLMA-T in different virtual worlds within VRChat: an indoor café (left) and an outdoor city (right).}
  \Description{This image consists of two side-by-side panels, both featuring ELLMA-T, the 3D cartoon avatar, in different immersive environments within VR. The avatar is the same as Figure 1.

Left Panel: The left panel shows ELLMA-T standing inside a cozy café environment. The café is designed with a rustic, brick interior, with a menu board displayed on the wall behind the counter listing various hot and cold drinks. To the left of ELLMA-T is a shelf stocked with baked goods and a wall clock hangs above it. The café ambiance is warm, with lighting that highlights the textures of the brick walls and wooden shelves. ELLMA-T is standing in front of the counter, facing the camera, and ready to assist the user in a conversational role-play scenario related to a café setting.

Right Panel: The right panel shows ELLMA-T in an outdoor environment resembling a European-style waterfront street. The scene features a stone bridge arching over a canal, with historic buildings lining the street and water visible in the background. The sky is clear and it is a sunny day. ELLMA-T stands on the cobblestone pathway near the bridge, facing the camera. This environment is for role-play scenarios where the user interacts with ELLMA-T in a street or outdoor setting, allowing for a variety of conversational topics related to urban exploration or travel.}
  \label{fig:different_worlds} 
\end{figure*}

\textit{Presenting Scenarios:} Upon knowing the user's proficiency level, ELLMA-T presents three different role-play scenarios focusing on various themes relevant to daily life. The user can choose the scenario that most interests them or suggest a specific scenario they'd like to practice. Once a scenario is chosen, ELLMA-T provides a detailed description of the scene, outlining the physical surroundings and significant objects. In VRChat, various virtual worlds replicate everyday settings like cafes, restaurants, outdoors, and supermarkets (see Fig. \ref{fig:different_worlds}). Once a scenario is selected, the user and ELLMA-T can be transported into a corresponding virtual environment. ELLMA-T prompts the user to complete tasks within the scene. For example, in a supermarket setting, the user is asked to find or describe specific items and practice the vocabulary learned.

\textit{Take Turns Conversations:} ELLMA-T and the learner engage in turn-taking conversations to complete the role-play scenario. Acting as the user's partner, the agent fosters natural and authentic dialogue within the chosen scenario. The conversation topic and task design follow the example of lesson plans of British Council English Teaching Lessons for adult learners \cite{british_council_lesson_2024}. In each scenario, ELLMA-T assumes a role such as a restaurant server, cashier, doctor, or friendly New Yorker on the street, allowing the user to conduct conversations that mimic real-life situations. 

\textit{Providing Feedback}: After the role-play, the agent provides feedback to the user, consisting of three parts. The first part offers general feedback that evaluates the user's performance concerning the difficulty level of the chosen scenario, highlighting one strength and one area for improvement. The second part, "Advice Moving Forward", provides practical tips on how to apply the concepts learned in the conversation to real-world situations. Lastly, ELLMA-T summarizes the key vocabulary, grammar, or sentence structures encountered during the interaction for the user’s future reference. ELLMA-T remains available to answer further questions or engage in additional practice sessions.

\subsubsection{Design Principle \#3: Communication Strategy and Back-channels}

EFL learners often encounter communication breakdowns due to insufficient vocabulary or incorrect grammar usage, which can negatively affect their self-confidence and motivation. Our system builds upon the communicative and scaffolding strategies by \cite{ayedoun_adding_2019} that may increase learner's willingness to speak in designing Embodied Conversational Agents (ECAs) for spoken English practice.

\textit{Mitigating Communication Breakdowns}: To address communication breakdowns, ELLMA-T provides encouragement to motivate the user to continue ("Come on, don't give up!"); or suggests phrases to help the user move forward in the conversation ("Maybe you can say..."). When the user's speech is unclear — whether due to incorrect vocabulary, grammar mistakes, or mispronunciation — ELLMA-T poses clarifying questions to prompt further dialogue ("You just said..., do you mean...?"). This approach fosters a supportive learning environment that encourages users to practice speaking without fear of immediate correction.

\subsection{System Implementation}\label{system_implementation}

\subsubsection{System Architecture}

\begin{figure*}[ht]
  \centering
  \includegraphics[width=\linewidth]{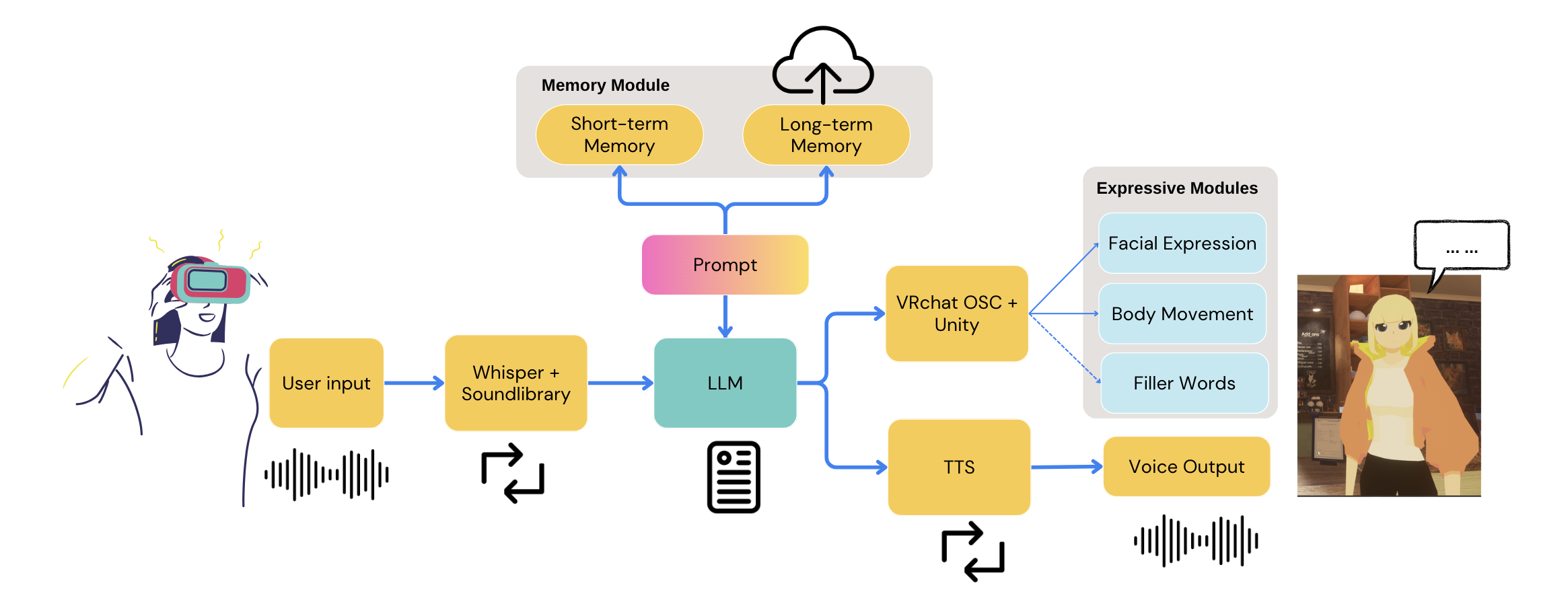} 
  \caption{System Architecture of the ELLMA-T. The architecture highlights the core components and data flow within the system.}
  \Description{System Architecture for LLM-Based Embodied Conversational Agent in VRchat.

This diagram illustrates the data flow between different system components enabling a conversational agent in VRchat.
User Input: Spoken input from the user is captured via a VR headset.
Whisper + Soundlibrary: Converts the audio input into text.
LLM (Large Language Model): Processes the text and generates a response, guided by a Prompt.
Memory Module: Includes Short-term and Long-term Memory to store interaction history for context.
TTS (Text-to-Speech): Converts the LLM-generated text into speech.
VRChat OSC + Unity: Manages the agent’s expressive behaviors, such as Facial Expressions, Body Movements, and Filler Words in VRchat.
Voice Output: The final speech output and expressions are delivered to the user.
}
  \label{fig:system_architecture}
\end{figure*}

The agent design in this project extended the work by \cite{wan_building_2024} 
which implements general ECA in VRChat and improves the response context-relevance by adding a \textit{memory} to the agent. Based on this implementation, we added a tutor workflow logic (see Fig. \ref{fig:conversation_tasks}) that controls the conversation flow. 

The learner controls an avatar in VRChat through a Quest 3 headset, able to make body movements and talk to ELLMA-T with the microphone. As illustrated in Fig. \ref{fig:system_architecture}, the audio input is converted into text by calling Whisper API \cite{openai_whisper_2024}, and then a text reply is generated from the GPT4 model by calling OpenAI API. The text response is converted back to Audio using the OpenAI TTS service (the 'alloy' voice) \cite{openai_tts_2024}. The user speech detection and silence detection are done by using Sounddevice Library \cite{python_sounddevice_play_2024} and Vosk \cite{python_vosk_vosk_2024}. We use a 2.0-second silence threshold to detect the end of the user's input, balancing the buffer time for users to pause and continue speaking with the overall wait time for the agent's response.

The system can also generate facial expressions and body gestures based on the user's input.
The LLM detects the user’s emotion by keyword detection and then mirrors that emotion by mapping corresponding body movements and facial expressions. The history of the current conversation session is stored in a list of JASON objects for the agent as the short-term memory and reference to generate more relevant conversations. The key information of the current conversation session is summarized by LLM and stored in a remote database. The back end of the agent is integrated into VRChat through VRChat OSC \cite{vrchat_osc_2024}. Additional animation for agent gestures and facial expressions can be added to Unity \cite{unity_unity_2024} and then incorporated into VRChat through VRChat OSC \cite{vrchat_osc_2024}.

\subsubsection{Multi-task Multi-turn Conversation System}
 
In designing and implementing a multi-task, multi-turn conversational agent, we encountered a challenge that necessitated a shift from a single-prompt approach to a more structured, multi-task approach. Our agent was tasked with managing an entire conversation involving multiple steps and decision points, for example, when to move from the introduction to the assessment task, how long the assessment should be, how to move on to the next topic, etc., within an interaction session. Initially, we experimented with a single-prompt approach, wherein the system prompt, we listed out all the steps and the criteria the agent should follow as system prompt in Appendix \ref{appendix:single_prompt}. However, we observed that the LLM struggled to follow the steps in the single prompt. The LLM often confused the sequence of actions, misinterpreted when one task should conclude, and when to transition to the next, or randomly restarted a certain task. This led to a breakdown in the logical flow of conversations, thus not ideal for users to interact with. This observation aligns with emerging findings in the field of LLM-based agents, where maintaining context and managing multi-turn interactions remain challenging, particularly in complex, multi-step tasks \cite{yi_survey_2024}. 

\begin{figure*}[ht]
  \centering
  \includegraphics[width=\linewidth]{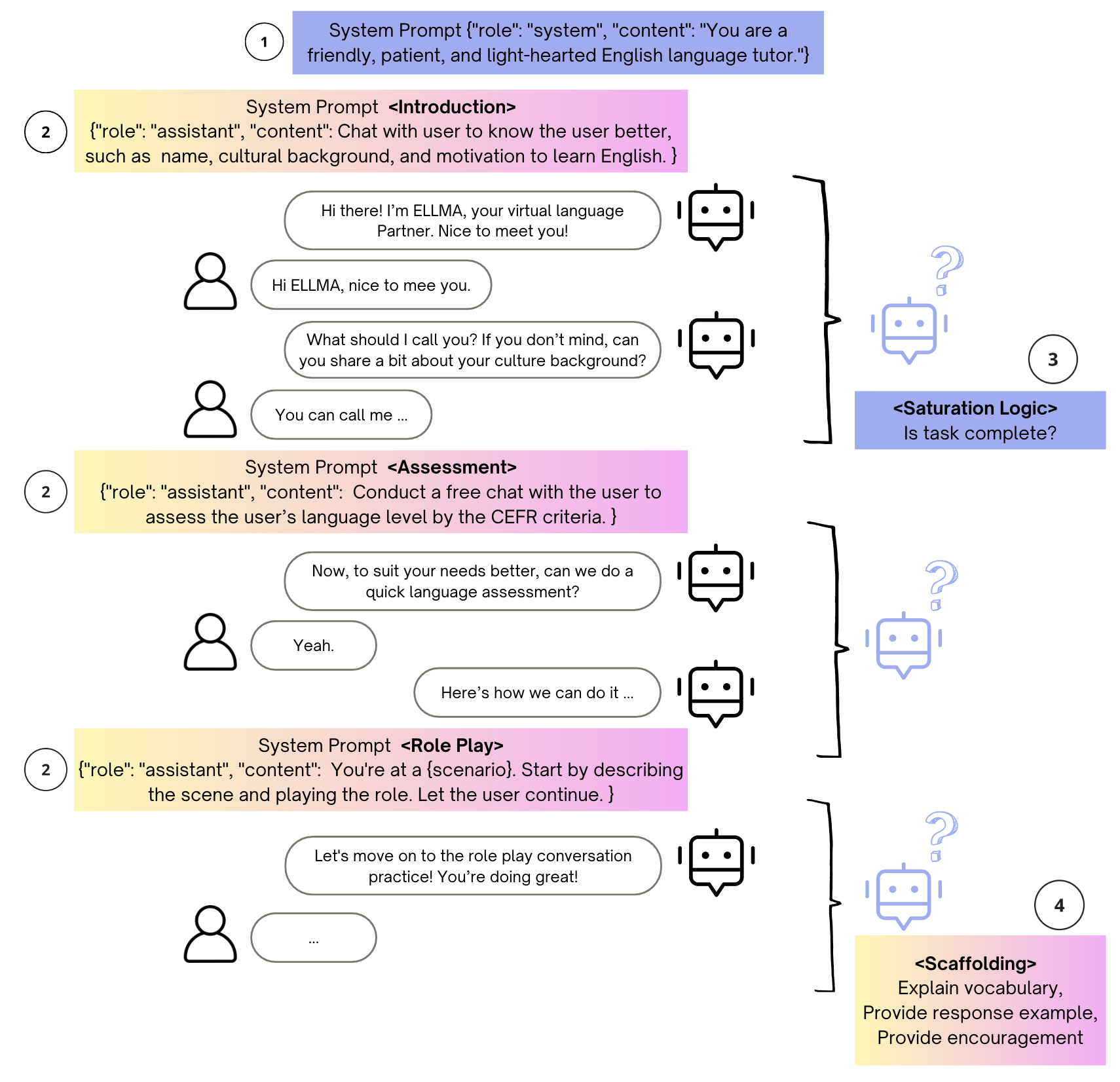} 
  \caption{Structure of Separate Prompts for Different Tasks. This diagram illustrates how prompts are structured and separated for various tasks within the system. 1) The system prompt establishes the agent's persona across all interactions. 2) Task-specific prompts guide the agent during the introduction, language assessment, role-play, and feedback. 3) A decision prompt helps the agent determine when to transition between tasks. 4) The prompt for providing scaffolding during role-play conversations.}
  \Description{This diagram illustrates the structure and flow of prompts used by the LLM-based agent to manage various conversational tasks:

System Prompt: Sets the agent’s overall persona as a friendly, patient, and light-hearted English tutor.
Task-Specific Prompts: Separate prompts are used for different tasks:
Introduction: The agent engages the user to learn about their background.
Assessment: The agent conducts a free chat to assess the user’s language level.
Role Play: The agent guides the user through a role-playing scenario.
During each task, there is a Saturation Logic that determines if the current task is complete and when to transition to the next task.
During the Role Play task, there is a prompt for Scaffolding that guides the LLM to offer explanations, examples, and encouragement as needed.

}
  \label{fig:prompt_structure} 
\end{figure*}

As an alternative, we adopted a multi-task implementation strategy (see Fig. \ref{fig:prompt_structure}. The core idea behind this approach is to decompose the overall task into smaller, simpler, and more manageable subtasks. Each subtask is then handled by a separately prompted instance of the agent (see Appendix \ref{appendix:multi-task sample promt}), allowing for greater control and reducing the cognitive load on the LLM at any given point. This strategy shifts the burden of maintaining deterministic logic from the LLM to the system designer, who predefines the sequence and structure of subtasks. By doing so, we ensure that the agent can navigate through complex interactions without losing track of the conversation's broader context. Each subtask is executed independently, with the LLM generating responses based on the specific context of that subtask. Although this introduces some variability in the conversation due to the inherent randomness of LLM-generated text, the overall structure remains intact, leading to more coherent and effective interactions. This approach is supported by recent developments in the multi-agent system for complex task dialogue systems, which suggests that breaking down complex tasks into modular components can significantly improve system performance \cite{wu_autogen_2023}.

\subsubsection{Prompt Engineering}
Effective prompt engineering is crucial for guiding LLMs like GPT-4 in performing specific tasks within a conversational system. While there are many ways to prompt LLMs \cite{schulhoff_prompt_2024}, there is currently no standardized workflow for developing the "perfect" prompt, as prompt design often involves a trial-and-error process \cite{zamfirescu-pereira_why_2023}. Our approach to prompt engineering was iterative and explored LLM's capabilities under zero-shot or few-shot conditions \cite{brown_language_2020} to facilitate language learning.

\section{User Study}

\subsection{Study Design}
To explore participants' experiences with ELLMA-T within the VRChat environment, we conducted a qualitative study using semi-structured interviews, and thematic analysis to analyze the data. The study aimed to explore the opportunities and challenges in LLM-based agent-supported language learning in a social VR setting. Our study focuses on overall interaction experience, preferences for system usage, and feedback for future improvements. This study was approved by the university's ethics board. 

\subsection{Participants and Setting}
For this preliminary study, we recruited 12 international graduate students (Table~\ref{tab:participant_info}) from [University Name removed for blind review] 
by posting recruitment ads via campus communication channels. Despite meeting the university’s rigorous English language standards, with a minimum IELTS score of 7.5, these students continue to encounter linguistic hurdles in their day-to-day lives and work environments. To that end, our participants were selected as a potentially representative sample of a population with mixed English proficiencies who aim to enhance oral proficiency in practical scenarios, such as social interactions and workplace communication.

\begin{table*}[th!]
  \caption{Participant Information}
  \label{tab:participant_info}
  \centering
  \small
  \begin{tabular}{@{}ccccc@{}}
    \toprule
    \textbf{Participant ID} & \textbf{Gender} & \textbf{Cultural Background} & \textbf{Language Proficiency} & \textbf{Motivation to Learn English} \\ 
    \midrule
    P1 & Male & China & Fluent & To improve language skills \\
    P2 & Male & India & Fluent & For work in Canada \\
    P3 & Male & China & Fluent & To enhance communication skills\\
    P4 & Female & China & Basic & To improve spoken English \\
    P5 & Female & China & Intermediate & For work and study in Canada \\
    P6 & Female & China & Intermediate & To enhance proficiency \\
    P7 & Female & China & Intermediate & To improve business English skills \\
    P8 & Male & Bangladesh & Very Fluent & To achieve advanced proficiency \\
    P9 & Female & Taiwan & Basic & For work and study in Canada \\
    P10 & Female & China & Basic & For job interview practice \\
    P11 & Male & China & Intermediate & To improve business English skills \\
    P12 & Male & China & Fluent & For work in Canada \\
    \bottomrule
  \end{tabular}
\end{table*}

\subsection{Procedure}
The study was conducted in person at the [University removed for blind review]. Upon arrival, participants received a brief overview of the study. They were informed that they would interact with an AI agent acting as an English language tutor using a Meta Quest 3 headset. For participants unfamiliar with VR, we provided instructions on how to wear the headset and navigate the VRChat world. The setup was pre-configured to ensure a seamless experience, allowing participants to bypass navigating menus or learning controls. Once equipped with the headset, participants were instructed to start recording their interaction with the ELLMA-T from their POV. The agent was overseen remotely by a researcher who also provided technical support. The interaction lasted \~15 minutes: introduction and assessment (5mins), role-play and feedback (7mins). We saved the conversation logs between each participant and ELLMA-T as CSV files for further analysis. Following the interaction, participants took part in a 30-minute semi-structured interview conducted by one of the authors experienced in qualitative methods. Interviews were audio-recorded with an iPhone 12 and transcribed in NVivo 14 for analysis. The interviews gathered qualitative data on participants' overall experience, comfort level, engagement, frequency of potential use, and preferences for using the system in different settings (e.g., group vs. individual; headset vs. desktop app). Follow-up questions were asked as necessary to probe deeper into participants' responses, particularly when initial answers did not fully address the intended topics or directions. The complete list of interview questions can be found in Appendix~\ref{appendix:interview_questions}.

\subsection{Analysis}\label{sec:analysis}
We employed an inductive (bottom-up) approach to analyze the interview transcripts, following the thematic analysis guidelines by Braun and Clarke \cite{braun_using_2006}. The analysis was conducted in several stages by three researchers, all authors of the paper. Initially, each researcher read over all the interview transcripts (step 1), and then independently analyzed a sub-set of the interview transcripts to identify preliminary codes, themes, and patterns, selecting relevant quotes and observations (step 2). The team then discussed and agreed upon a final set of codes through multiple rounds of coding (step 3). To ensure the robustness of our analysis, the interview transcripts were cross-referenced with video recordings and notes taken during the interactions (step 4). Finally, all themes were discussed and defined by three researchers/coders (step 5) and written up (step 6).

\section{Findings}
\label{sec:findings}

In this section, we describe four key themes: agent perception and its role in social VR, language learning in social VR, task design and perceptions of LLMs efficient in language learning, and interest in continued usage.

\subsection{Agent Perception and its Role in Social VR}
\label{subsec:agent_perception_and_role}

\subsubsection{ELLMA-T Human-like qualities} 
\label{subsubsec:human_like_qualities}
Participants compared their experience with ELLMA-T to that of conversing with a real human tutor. Participants frequently commented on the agent’s human-like qualities, noting how the interaction felt like conversing with a real person rather than an AI. As P4 expressed, \textit{"I love the way it talk(s) to me. It (is) just like a real teacher; the interaction is really good."} Participants highlighted that the conversation flowed naturally, demonstrating ELLMA-T's ability to listen, understand, and respond appropriately. For example, P8 appreciated that the agent \textit{"was listening to me,"} suggesting that it responded in a way that related to their personal experience. P2 also observed, \qq{"First of all, it started off with the cultural experience. So then the moment I mentioned that I'm from India, the greeting changed from 'Hello' to 'Namaste,' which is like 'Hello' in Hindi in our language. So that 'Namaste,' that added a personal touch. Then it tried to understand how you feel in day-to-day conversations."} This ability to personalize the interaction and subsequent conversation flow led P2 to conclude, \textit{"It's like... I'm talking to someone in person, so it's not an AI for me right now."}

\subsubsection{Importance of Adaptability and Role Switching Depending on Personal Preferences and Social Context}
\label{subsubsec:agent_role}

Participants expressed varied preferences regarding ELLMA-T’s role in facilitating language learning. Some preferred the agent to take on a traditional teacher role, similar to that of a classroom setting. For instance, P1 suggested, \textit{"The agent could be like the teacher, and then the other people would be students. I kind of still feel like that — be more traditional, the agent as the teacher and the head of a classroom."} Others envisioned the agent in a more supportive role, acting as a facilitator or social glue in multi-person interactions. P5 imagined this role as helping to bridge gaps in conversations: \textit{"And also they can fill the gap. Okay, what if two people are introverts and they probably don't have much in common to talk about, the AI agent could ask questions or lead the conversation."} Additionally, some participants saw the agent as a feedback provider during real-human conversation practices. As P6 noted, \textit{"You can have a conversation with a real person and learn from what they said, with the feedback provided by the agent."}

\subsubsection{Perceptions of Agent Tone, Appearance, and Behaviour}
\label{subsubsec:tone_appearance_bodymovement}
We observed how nuanced perceptions of the agent's tone, physical appearance, and behaviour impacted the overall interaction experience. 

\noindent
\textbf{Agent Tone:} Some participants noticed that the agent's language tone when performing different tasks influenced their learning experience and how they perceived the agent. For example, P6 noted a clear distinction between the agent's tone during the role-play and when giving feedback: \textit{"I would hope that the (role-play) conversation is more casual, but I expect the feedback to be professional."} P8 highlighted the importance of providing encouraging phrases combined with an appropriate tone: \qq{"Right now it's very robotic, and so it just gives me basic feedback of ... great improvement, but I liked certain things that you did... she did mention those, but it seemed very... like I was just typing into a computer and it's coming out."} P8 also added that the lack of inflection and tone variation in different contexts can make the agent's interactions seem less engaging and less human-like: \textit{"If you're having a conversation, there are nuances on how we speak and certain sounds...The ups and downs in the way we speak."}

\noindent
\textbf{Agent Appearance:} Some participants appreciated the neutral appearance, as it did not distract from learning or convey judgment. P1 remarked, \textit{"The tone overall is general. It's not too aggressive or too... General is good; it didn't create a weird or odd atmosphere."} 
However, others preferred more personality for fun and engagement. P10 suggested, \textit{"Maybe we can just use some famous characters from games, literature, or anything, and we can set up the environment ... and it will behave and speak like that."} Several participants also expressed a preference for customizing the agent's appearance. Suggestions for different avatars included realistic (P11, P12), cartoonish (P3, P10), or non-human-like animals (P6), as well as the ability to modify attributes like hair color, skin color, body type, gender (P6), and clothing (P4).

\noindent
\textbf{Agent Behaviour:} Similar to preferences for tone and appearance, P7 acknowledged the minimal movement of the agent as beneficial for avoiding distractions from learning English. \textit{"I didn't really look at how he moved... I feel like maybe if add on too many moves, it can be distracting from learning English."} (P7) However, others suggested that natural gestures, such as head nodding and eye contact, would enhance engagement and create a more lifelike interaction. P11 stated, \textit{"It should be representative of how people think of themselves. And then the blinking, the head nodding, it should all be there inside the VR environment."} Additionally, participants highlighted the importance of cultural and contextual factors, such as the need for consistency between the agent's voice and visual identity and the potential for scenario-based adaptations of the agent's personality and appearance. 
P12 mentioned \textit{"I think for different scenarios, we can have changes in personality or change the avatar itself to simulate different environments and how different people speak. That would make a difference, I think."}

\subsubsection{Freeing Interactions of Social Norms that Do Not Apply to AI}
\label{subsubsec: agent_approachable}
All participants found the agent approachable and comfortable to interact with. One significant advantage noted was the absence of social pressure or expectations tied to cultural or social norms. P5 highlighted this, stating, \qq{"You talk with a person, and if you don't properly tell them certain things, you might hurt their feelings. But if an AI gives you feedback, it's like, oh, a machine is telling you the feedback, so it's okay. They don't have those thoughts like, oh, what if I hurt this person's feelings?"}

\subsection{Language Learning in Social VR}
\label{subsec:language_learning_in_social_vr}

\subsubsection{Situated Learning Supported by VR Affordances} 
\label{subsubsec:environmental_context} 
Participants spoke about how their language learning was supported through VR's unique affordances: 3D interactivity with objects and avatars, contextualized learning in an immersive environment, and enhanced emotional support through an embodied agent. 

\noindent
\textbf{Interactive Learning:}
Participants described how being in a VR environment added significant value to their language learning experience. The interactive nature of VR was frequently highlighted, with P1 noting, \textit{"The VR added the interactions. What I need to do is just speak. I don't have to type."} The environmental context provided by VR, along with the embodied agent, created a sense of immersion that made the interaction feel ‘real.’ 
As P5 shared, \textit{"I feel like I'm talking to a kind of a real person standing in front of me. I mean, it’s VR after all. So that's a huge difference."}

\noindent
\textbf{Contextualized Learning:}
The VR environment provided valuable context for language learning, helping participants understand how to use language in real-life situations. P8 mentioned: \textit{"For example, the bakery, it'll be in that place. We walk by it, and it was like, oh, this is the bakery, and if you come here, then you can get these types of soups. But it'll be applicable to my real-life scenarios as well."} Similarly, P12 emphasized the advantage of having a virtual environment that corresponds to the learning task: \textit{"I think if we can actually have a coffee shelf, where there are espresso machines and everything. I think having an actual environment will make the difference, and that's the advantage of VR."}

\noindent
\textbf{Emotional Support:}
Participants also discussed the emotional connection they felt with the embodied agent in VR. P1 noted that VR added an emotional dimension to the interaction: \textit{"It adds emotions. So having a static or 2D agent just living on a screen, it feels like there's less emotion attached to it."} Other participants mentioned how adding personality to the agent in VR increased the sense of emotional connection. P7 suggested, \textit{"Or maybe they [people using this system] need more emotional support. So that may be a good way to give the agent a personality."} P9 and P11 emphasized the importance of adding physical reactions and body language, with P9 stating, \textit{"Maybe more physical reaction. Like somebody I really talk to you and looking into your eyes,"}.

\subsubsection{ One-on-One interactions alleviate Speaking Anxiety, Group Interactions support Shared Learning Experience}
\label{subsubsec:private_or_group}
Participants expressed varying preferences for interacting with the agent in VR, either in a private space or in a group setting. Participants felt they would experience anxiety speaking in front of other people and preferred one-on-one interaction. As P2 articulated, \textit{"I've had this problem, and I know many of my friends also have this problem. They kind of get, when they don't have their confidence over the language, they kind of get overwhelmed and a little intimidated by being in the group."} 
Some participants preferred one-on-one interaction because they valued the direct interaction and feedback from the agent. As P10 noted, \textit{"It's one-on-one, so it can assess me and it can target my weakness."}

On the other hand, others preferred group interactions, recognizing the additional benefits of more engagement, peer support, and shared learning experiences. VRChat, as a social platform, offers the flexibility of choosing avatars and social modes, which can enhance learning with peer learners. P10 suggested, \textit{"If we as learners can interact with each other in VR, that would be fun because we won't care about what each other looks like or where we are, and we can focus on practicing."} 
Moreover, P7 highlighted the potential real-life benefits of group interactions in VR, such as preparing for group meetings and work environments, which cannot be simulated by interaction with the agent alone: \qq{"I would love to have a group setting because it might be helpful if you're going to work or have a group discussion. In real life, you have to catch up with what different people are talking about. So it might be more challenging, but it can be fun too."}

\subsection{Task Design and Perceptions of LLM's Efficacy in Language Learning Tasks}
\label{subsec:llm_efficacy}
The potential effectiveness of the learning experience was a central concern of participants for the design of the system, particularly in terms of task design relevance, assessment accuracy, conversation content, feedback, and the support provided by the agent. 

\subsubsection{Task Design Relevance}
\label{subsubsec:task_design}
Our analysis showed that the system’s ability to address the contextual needs of learning spoken English for adult language learners in English-speaking countries was a recurring theme. Participants acknowledged that the agent effectively targeted specific challenges they faced. One significant challenge highlighted was the lack of opportunities for learning or receiving feedback even though they are already in the language environment, as P5 explained: \qq{"Usually when you practice your English, it's just... well, I'm studying here already in a foreign country. No one will give me their feedback. They can't, I think. I do know this is Western culture. People give more positive feedback. But this one, it gave me more reality (realistic) feedback. You don't get it in real world."} Another participant, P5, emphasized the gap in informal conversation learning, which is often missing from formal and academic settings: \qq{"Something I really want to learn about is how to use English when you try to agree with someone's idea. It can be super, super simple for a native speaker because they don't really know, oh, this is hard for a non-native speaker. I want to respond. When someone says something, I can say something like, 'oh', 'amazing', 'absolutely'—that kind of simple words. It is something that can be super hard for me."}

Participants valued the form of role-play conversation for its ability to provide a real-life conversation experience. As P7 noted, \textit{"Because the conversation and dialogue is about daily life communications where you can learn something that is useful for you, useful for you to learn English."} 

Participants also suggested specific scenarios they would like to practice, such as renting an apartment as an international student (P10) or participating in a mock interview (P6).

\subsubsection{Assessing the Learner's Language Ability}
\label{subsubsec:assessment}
The majority of participants found the agent's assessment of their language abilities to be accurate and consistent with their self-perceived levels. Five participants described the assessment as "seemed appropriate" (P11), "accurate" (P3, P8), "very accurate" (P4), and "fit to my level" (P9). Even if the assessment results were not completely precise, the respondents' feedback suggests that the agent's assessment met their expectations. P7, while not directly commenting on the accuracy of the assessment, found it useful as a form of descriptive feedback: \textit{"She actually gave me a level, like a B1, which is quite interesting. I feel like I also want to know, oh, what kind of level I'm at and she gave some description like 'you can describe your feeling', which is a good thing."}

An interesting case arose with P10, who chose not to follow the agent's instructions and remained silent during the assessment. Consequently, the agent categorized her CEFR level as A1 (basic) \cite{coe_cefr_2024}, a result she strongly disagreed with: \textit{"It has an agenda, so it wants me to repeat after it. I don't want to do it... (The assessment was) not very accurate. It says I'm A1 level, but at least I think I'm C1 or C2."}

\subsubsection{Role Play Conversation Content}
\label{subsubsec:role_play_content}
The role-play conversations provided by the agent were well-received by participants. ELLMA-T showed creativity and adaptability in generating conversations, often performing well in various roles without requiring extra prompts. P3 and P8 both described their role-play experiences as "surprisingly good." For example, during a painting gallery scene where the participant played the role of a salesperson and the agent played the customer (role-play conversation 1, see Table \ref{tab:roleplay_example}), P3 found the experience highly engaging and realistic: \qq{ "It provides very practical scenarios to have a conversation. And also this tutor can reflect according to my answer, because it requires abstract painting. And I say, I can give you a portrait. And they (the agent) say(s), I don't need a portrait. I need (a) more abstract one. So that's a good point I think, because you can feel that you really talk with a person. A person can give you (a) correct reflection. That is wonderful."}

This example highlights the LLM's ability to leverage its vast pre-trained knowledge to create vivid, realistic and immersive conversations, moving beyond the dry, textbook-like exchanges that often characterize traditional language learning. The agent’s ability to add nuanced details based on the role and situation enhances the realism of the interactions. As P8 noted, \textit{"Yeah. It's not just that kind of dry back and forth... Yeah, it's not like an IELTS exam where you just ask a question and get an answer."} This dynamic approach brings the context of the conversation to life, making the learning experience more engaging and effective.

\begin{table*}[th!]
  \caption{Example of Role-Play Conversation in Painting Gallery Scene}
  \label{tab:roleplay_example}
  \centering
  
  \begin{quote}
  ELLMA-T [setting the context for role-play conversations]: So let's imagine we are in an art gallery filled with beautiful and eclectic pieces of art. You are a knowledgeable and enthusiastic art curator and I'm an interested buyer exploring the gallery, maybe looking for something to spruce up my home. The conversation could start with me walking into the gallery. Okay, here we go.
  \end{quote}

  \vspace{1em} 

  \begin{tabular}{@{}clp{8cm}@{}}
    \toprule
    \textbf{Event Line} & \textbf{Interaction Context} & \textbf{Excerpt} \\ 
    \midrule
    1 & Agent (Customer) & Hello there. This is my first time visiting this gallery. Could you show me a few standout pieces and tell me their stories?\\ 
    2 & User (Salesperson) & Yeah. Can you first tell me what kind of artwork you want to buy or you're looking for? \\ 
    3 & Agent (Customer) & I'm very interested in contemporary paintings. Preferably something vibrant and abstract. I'm looking for a piece that could be the focal point of my living room. It's a fairly large space with neutral colors.\\ 
    4 & User (Salesperson) & Can we look at this portrait? This portrait is about someone who is in the renascence age and in this portrait, you can say the background is in the dark, broad color. And you can say the portrait is very clear and the face of the person in the portrait is very dynamic and active. And what do you think about this?\\ 
    5 & Agent (Customer) & The contrast between the dark background and the vivid facial expression must create a fantastic visual impact. However, I am more interested in abstract art rather than portraits. Do you have any abstract pieces with bold, colorful strokes? \\ 
    6 & User (Salesperson) & Yeah. If you are looking for some abstract. How about the Impressionism?\\ 
    \bottomrule
  \end{tabular}
\end{table*}

\subsubsection{Feedback Effectiveness}
\label{subsubsec:feedback}

 Participants frequently commented on the agent’s feedback as one of its most valuable features. For example, P2 appreciated the specific feedback on tense usage, stating, \textit{"So with my tenses, if there was a mistake with the tenses, that's there. What can I improve, that's there. The feedback was absolutely correct. Even I know that I goof up in my tenses sometimes."} 

 P7 also emphasized the benefit of receiving feedback on common errors made by non-native speakers: \textit{"I feel like it improved a lot because usually, as non-native speakers, we can't really tell when we're using the wrong tense or wrong words, but they mention it as feedback. So I feel like it's really helpful."}

Despite recognizing the value of the feedback, participants also expressed a desire for more personalized and detailed responses. They suggested that the feedback could be tailored based on their learning history and user profile, with specific examples of mistakes and corrections. P11 pointed out that, \textit{"For learning English, I think that AI isn't giving very specific responses. It's giving very general answers. If a person was actually normally listening, the tutor would reply with an example of what I said wrong and then rephrase it and give an example of what it should be."}

Additionally, while one participant found the feedback "educational," they also mentioned that it could be too lengthy: \textit{"I think the feedback is sometimes too long for me because it will illustrate a lot of feedback evaluation."} This suggests that while comprehensive feedback is appreciated, there is a balance to be struck between thoroughness and conciseness to maintain user engagement.

\subsubsection{Scaffolding Effectiveness}
\label{subsubsec:scaffolding}
The agent provided various forms of support and encouragement throughout the sessions. Verbal cues like "You are doing great!" and "Good job doing [activity]..." were consistently used to motivate participants. When participants struggled to generate dialogue during role-play, the agent offered scaffolding by presenting examples for them to repeat. As P7 noted, \textit{"I do like the part that he actually gave some hint, 'for example ...', 'Oh, you can say that ...'."} Additionally, the agent asked clarification questions when the meaning of the user's input was unclear and provided explanations when participants were unsure how to proceed. P1 described an instance where:
\textit{" She asked me to straight up tell her the story about something and I didn't quite catch that the first time. I asked her what are the options that I have. And then she gave one and I liked it."}

An interesting observation was made by P12, who found that despite having an accent, the agent accurately captured his input: \textit{" I think it captures words better than I thought it would. Maybe I have an accent sometimes. I didn't expect it to be this good."} This highlights a common communication challenge among English language learners, which presents a significant design challenge for the system.

\subsection{Interest in Continued Usage}
\label{subsec:continued_use}
\subsubsection{Likelihood of Continued Usage}
\label{subsubsec:likelyhood_of_continued_use}
 Nearly all participants indicated that they would use the system regularly to practice their language skills, with usage times ranging from 30 minutes to four hours daily (P1, P2, P3, P6, P10, P12), or one to two times per week (P7, P8). Some participants noted that they would increase their usage if the physical fatigue associated with heavy headsets could be mitigated (P4, P8). Additionally, some participants indicated that they would increase their usage when preparing for specific scenarios such as job interviews (P9). These findings suggest our system has strong potential for supporting sustained language learning over time.

\subsubsection{Factors Negatively Impacting Continued Usage}
\label{subsubsec:negative_factors_continued_use}
Despite the positive indications for regular use, we identified several challenges related to task design and conversation flow, which could cumulatively and negatively impact long-term usage. One participant highlighted how the response latency could become more problematic over time, noting, \textit{"So this seems to be okay, but as the speed varies, the transaction time — so the time that I need to spend with the AI — will increase more as per the response time"} (P2). Additionally, breakdowns in conversation flow were noted when the agent responded too quickly or occasionally interrupted the participant, further detracting from the overall experience. Participants also reported issues with task design, such as unclear task explanations or a lack of concise instructions when switching tasks. 

\phantomsection
\noindent
\textbf{Importance of Short and Concise Instructions for Learning:}
\label{customsubsec:context_setup}
During all the learning sessions, participants generally followed ELLMA-T's instructions. One participant commented that the instructions were clear, stating, \textit{"All the rules and regulations, all the prompts that the user needs to follow are there"} (P2). However, not all participants shared this experience. One participant expressed panic and confusion when starting the session, noting, \textit{"I feel like just having a short, precise description of the context would help. I just couldn't figure out what was going on"} (P1). An observation from the video recordings supports this concern; ELLMA-T sometimes provided instructions that were too lengthy to fit within the chat window, making them difficult for participants to follow. When this occurred, participants often had to ask ELLMA-T to restate the task, which disrupted the flow of the session. 

\phantomsection
\noindent
\textbf{Latency-related Conversation Flow Breaks:} 
\label{customsubsec:latency}

Latency was an issue in generating responses commonly associated with currently available LLMs. Participants frequently mentioned that the delay in response time made it clear they were interacting with a machine rather than a human. As one participant noted, \textit{"The response time is what will make you... okay, you are talking to a machine right now"} (P2, P5). Another echoed this sentiment, \textit{"I think his feedback is a little bit slower. Sometimes I need to wait for his response"} (P4). This delay in processing left participants feeling disconnected from the conversation, as highlighted by P10: \textit{"It doesn't feel like a real conversation. There are gaps and silent time."}

\phantomsection
\noindent
\textbf{Conversation-pacing-related Flow Breaks:} 
\label{customsubsec:pacing}
Another major disruption came from the rigid turn-taking structure of the conversation. Unlike human interactions, where there may be overlaps of speaking or long pauses while turn-taking, ELLMA-T strictly adhered to turn-taking by waiting for the user to finish speaking before responding, and did not respond to interruptions. This often led to interruptions and a lack of responsiveness. For example, two participants described how "There were a few times when I intended to speak, but she just continued" (P1), "I noticed I couldn't interrupt the agent. Sometimes I find it hard to tell whether it's my turn to speak." (P12). Others expressed frustration with the agent's inability to adapt to their conversational pace, with P7 noting, "She doesn't wait for me to speak".
Addressing these issues will be essential for improving the overall user experience in long-term language learning and making the agent feel more like a genuine conversational partner.

\section{Discussion}
\label{sec:discussion}
To address our research question, we contextualize our findings with prior literature and discuss the potential of embodied LLM-based agents for language learning in social VR 
in the following sections. We first discuss the key qualities reported by participants of ELLMA-T to support English language learning; second, the advantages of situated language learning in social VR; and third, the benefits and challenges of embodied LLM-based agents for language learning in social VR.


\subsection{User Experience of ELLMA-T for Supporting English Language Learning}
\label{subsec:benefits_potentials_of_agent}

Overall, our findings indicate that ELLMA-T possesses several key qualities that align with the existing literature's criteria for an effective language tutor: human-like interaction, enhanced emotional connection through embodiment, a demeanour that is accessible and approachable, and provision of encouragement and support. 

\subsubsection{Emulating Human-Like Interaction} One of the key strengths of ELLMA-T is its ability to emulate human-like interaction, a quality that participants frequently noted in Section \ref{subsubsec:human_like_qualities}. The perception of the agent as 'human-like' suggests ELLMA-T's success in resembling a human tutor's ability to maintain an interactive and meaningful conversation. This finding supports previous research that underscores the significance of ECAs mimicking human-to-human interactions and highlights the critical role of 'believable' interactions in conversation system design with ECAs \cite{cassell_embodied_2000}. 
As our findings in Section \ref{subsubsec:agent_role} indicate, ELLMA-T is seen in the roles typically fulfilled by humans such as a tutor or a buddy. Specifically, participants appreciated how the agent could listen, understand, and respond appropriately, making the learning experience feel more natural and less mechanical.

\subsubsection{Enhancing Agent-Learner Emotional Connections} Existing literature suggests that the embodiment of an agent in language learning enhances the immersive experience and strengthens the emotional connection between the agent and the learner, which positively impacts learning outcomes \cite{lefkowitz_second_2009}. Our findings in section \ref{subsubsec:environmental_context} support this, as participants reported feeling a stronger emotional connection with a 3D embodied agent in the VR environment (P1, P6). Furthermore, the data in section \ref{subsubsec:tone_appearance_bodymovement} suggests that adding personality, personalized tone, appearance, and body movement to the ECA can further enhance this emotional connection (P7, P9, P11).
However, previous research presents mixed views on the impact of incorporating ECAs in a VR environment for learning, as it may increase cognitive load for learners \cite{petersen_pedagogical_2021}. This also aligns with our findings that some participants preferred the agent to have a neutral tone, minimal facial expressions, and body movements, as these were perceived as "less distracting" to learning (P1, P7).

\subsubsection{Importance of Non-Verbal Cues} Emotional support is recognized as a critical factor for sustained language learning \cite{dincer_common_2013}. In our implementation of ELLMA-T, we adopted communicative and affective strategies based on \cite{ayedoun_adding_2019} to increase learners' willingness to communicate (WTC) by providing verbal encouragement and support that was appreciated by the participants (see Section \ref{subsubsec:scaffolding}). However, our findings also indicate that verbal cues alone were insufficient, particularly in the interactive VR environment (see Section \ref{subsubsec:tone_appearance_bodymovement}). This finding aligns with existing research, which suggests that non-verbal cues significantly enhance the perception of human likeness and the effectiveness of communicative tasks with ECAs \cite{cassell_may_2000}. Participants reported that the agent's voice inflections influenced their learning experience, with some noting that a neutral tone sometimes felt slightly negative or lacking in encouragement, even when verbal encouragement was present. This echoes the result of a recent study suggesting that ECAs with extrovert personalities, complemented by expressive facial movements and body language, are perceived as more effective for general learning purposes \cite{sonlu_effects_2024}.  However, in the context of language learning in social VR, our findings in Section \ref{subsubsec:tone_appearance_bodymovement} suggest that the need for nuanced interactions becomes even more critical. Participants expected verbal cues to be complemented by appropriate tones, facial expressions, and body movements to be fully effective. In addition, these non-verbal cues need to be appropriate for the language learning scenario and social context.  Our observation also indicates that the effectiveness of the agent’s encouragement was also less noticeable, likely due to technical difficulties that interrupted conversation flow and the generally high language proficiency of the participants, which may have reduced the perceived need for support. This points to a broader challenge in designing ECAs with various personality and non-verbal cue combinations for learners with different language levels.

\subsubsection{Reducing Speaking Anxiety} As findings in Section \ref{subsubsec:private_or_group} suggest, ELLMA-T shows the potential of reducing speaking anxiety among participants (P8), especially in one-on-one interactions. Participants felt more at ease when interacting with ELLMA-T in the VR environment, shielded by an avatar and free from the social pressures typically associated with human-to-human interactions that do not necessarily apply to human-AI interactions (see Section \ref{subsubsec: agent_approachable}). The expectation of interacting with an AI, rather than a human, naturally reduced the pressure often associated with speaking in a foreign language. This finding supports the effectiveness of ELLMA-T in language learning, which aligns with existing research that emphasizes the importance of reducing anxiety in language learning environments \cite{woodrow_anxiety_2006}, especially for adult immigrant learners who often experience heightened pressure in social situations \cite{adjagbodjou_envisioning_2024}. ELLMA-T's perceived approachability and the fact that it is always available and accessible further contributed to this reduced anxiety, positioning the system as a valuable tool for continuous language practice without the social constraints typically associated with human interactions.

\subsection{Situated Language Learning in Social VR}
\label{subsec:discussion_social_vr}
The situated learning theory explains how learning is most effective when it occurs within the relevant context\cite{lave_situated_1991,brown_situated_1989}, our findings in Section \ref{subsec:language_learning_in_social_vr} correspond to this theory. Participants expect to have the VR scene, and characters with different personas in VR environment corresponding to the role-play conversations to have a more realistic experience of a potential real-life situation in which use of learned language may be important. In addition to the advantages the VR environment provides to learning, the opportunities with Social VR are also discussed in the following sections.

\subsubsection{Flexibility in Interaction Modes}
The findings in Section \ref{subsubsec:private_or_group} suggest that a notable advantage of using social VR is the possibilities it offers in switching between private interactions with the agent and agent-facilitated group interactions. Participants appreciated the ability to engage in one-on-one sessions with ELLMA-T, which helped reduce speaking anxiety (see Section \ref{subsubsec: agent_approachable}) and allowed for personalized feedback. At the same time, VRChat, as a social platform, also has the potential to meet learners’ social needs and foster a sense of community and group affirmation as discussed in Section~\ref{subsec:discussion_social_vr}. In addition, the option to transition to group interactions facilitated by the agent introduces a dynamic that is difficult to replicate in traditional language learning settings. Previous research has shown promising results for human teacher-facilitated collaborative language learning in VR between human students \cite{jackson_collaboration_2000,peterson_efl_2012,cantone_contextualized_2023}. With ELLMA-T potentially facilitating similar group interactions, it provides a promising direction for further exploration. 

\subsubsection{Trade-offs Between VR headsets and 2D Desktop Interfaces}
VRChat provides users with the flexibility to interact with the platform through VR headsets or a 2D desktop interface. While the immersive qualities of VR provide significant benefits, our findings in Section~\ref{subsubsec:likelyhood_of_continued_use} also highlight the trade-offs associated with prolonged use of VR headsets. As noted in previous research, the availability of the technology, cognitive load and fatigue associated with VR are significant barriers to widespread use \cite{petersen_pedagogical_2021}. This issue was echoed by participants in our study, with some participants expressing a preference for interacting with a 2D desktop agent due to the convenience and reduced physical strain. This suggests that while VR offers powerful immersive experiences, a more accessible 2D interface could broaden the system’s appeal and usability, especially in longitudinal studies where extended use is necessary.

\subsection{LLMs for Language Learning: Performance and Potential}
\label{subsec:discussion_llm}
Findings from Section \ref{subsec:llm_efficacy} suggest that ELLMA-T, powered by GPT-4, performed well on current teaching tasks, particularly in generating role-play conversations and providing feedback. This aligns with recent research \cite{caines_application_2023}, which confirms the LLM's capability in learning material generation and assessment. The ability of LLMs to create flexible, context-sensitive content presents significant advantages over previous scripted or NLP-assisted systems, which often suffered from communication breakdowns due to technical limitations \cite{zhang_justspeak_2021}. However, present limitations of LLMs such as latency or lack of precise turn-taking logic can lead to breakdowns in communication flow that may impact users' motivation to use the system for prolonged time (see Section \ref{subsubsec:negative_factors_continued_use}).

\subsubsection{ELLMA-T as a Personalized and Creative Tutor}
The flexibility and creativity LLM brings to role-play conversations are central to the human-like qualities of ELLMA-T as discussed in Section \ref{subsubsec:role_play_content}. LLMs can also generate dynamic and coherent responses \cite{yang_harnessing_2024} that can adapt to the learner's input in real-time \cite{caines_application_2023}. 
However, to fully exploit the potential of LLMs for different language learning tasks, more intricate prompting and contextual understanding are necessary. Our findings in Sections \ref{subsubsec:tone_appearance_bodymovement} and \ref{subsubsec:feedback} support prior work that indicates that the quality of the LLM’s output, particularly the wording of the output, should adapt to the context in which it is used \cite{barambones2024chatgpt}. We hypothesize that improving the LLM's performance will require: 1) a better understanding of conversational context, for instance, the agent's role and the expected tone; 2) developing more sophisticated techniques to generate appropriate responses in various contexts. Additionally, future work can explore how text responses can be effectively mapped to corresponding voice tones, facial expressions, and body movements under different contexts.

\subsubsection{Level Assessment and Difficulty Level Adjustment} The results from 
Section~\ref{subsubsec:assessment} indicates that the GPT-4 assessment of participants' language level was overall accurate and in agreement with the findings by \cite{caines_application_2023}. However, the assessment format could be better designed with clearer instructions and a more flexible format given the feedback from participants who do not want to follow a fixed agenda. Following the assessment, the findings in Section~\ref{subsubsec:assessment} infer that GPT-4 likely adjusted the language difficulty level for subsequent conversations based on the results. This inference is supported by feedback from P3 and P8, who were assessed at a high proficiency level (C1) and mentioned that the vocabulary used in their subsequent conversations might be too difficult for beginners. This suggests that the system correctly matched the difficulty level to their advanced assessment. However, we have not explicitly assessed how well GPT-4 used in ELLMA-T was adjusting task difficulty based on the previous assessment of the learner's language proficiency. Given that no other participants raised concerns about the difficulty level, it is possible that the adjustments were suitable for their assessed levels, but without a more comprehensive assessment, this remains speculative. Future studies can include specific questions about participants' perceived difficulty level and linguistic metrics, such as word frequency, sentence length, and syntactic complexity, to better understand how well LLMs can tailor language task difficulty to individual needs.

\subsubsection{Opportunities and Challenges of Designing LLM-based ECAs for Informal Spoken English Learning in Social VR}
\cite{adjagbodjou_envisioning_2024} emphasized the importance of informal language practice as distinct from formal academic learning, highlighting a gap in tools designed specifically for this context. Our study echoes this view, with findings from Section~\ref{subsubsec:task_design} showing that participants valued the opportunity to practice speaking English in real-life, informal settings. Despite being in an English-speaking environment, participants in our study reported challenges such as a lack of feedback and opportunities for language improvement in Section~\ref{subsubsec:task_design}, underscoring the need for effective feedback mechanisms. Common challenges faced by EFL learners, such as anxiety and the need for support to increase WTC \cite{ayedoun_adding_2019}, were partially addressed by our system (see Sections \ref{subsubsec:private_or_group}, \ref{subsubsec: agent_approachable}, and \ref{subsubsec:scaffolding}). However, the challenge of designing effective verbal and non-verbal cues, as discussed in Section~\ref{subsec:benefits_potentials_of_agent}, poses challenges in ensuring the system can accurately interpret and respond to users' intentions. Additionally, communication obstacles common among EFL learners — such as accents, grammar mistakes, incorrect expressions, low speaking volume due to lack of confidence, and intermittent speaking due to unfamiliarity with the language \cite{adjagbodjou_envisioning_2024} — further complicate the system's ability to process and respond effectively. These issues are more challenging for an audio detection system to interpret compared to inputs from native speakers, highlighting the need for more advanced solutions tailored to the specific needs of EFL learners.

\section{Design Implications}
\label{sec:design_implications}

We present five design implications for LLM-based ECAs to facilitate spoken English learning in Social VR based on our key findings.

\subsection{Promoting Personalized and Adaptive Learning through Design}


\textit{Design Implication 1: Design LLM agents with contextual and cultural relevance in the adaptive learning experience based on user information, user preference, and learning history to support personalized and adaptive learning. }

The analysis in Sections \ref{subsec:agent_perception_and_role} and \ref{subsec:discussion_llm} highlight the importance of adaptive learning experiences in language learning tools. Participants wanted responses tailored to their proficiency levels, challenges, and progress. LLMs have the potential to suggest learning materials based on user information, user preference, and learning history \cite{darsih_learner-centered_2018}. However, this requires the agent to have a more sophisticated memory design and advanced reasoning ability. Emerging studies have focused on developing different memory mechanisms for increasing context-awareness and adaptation to user behaviour patterns. For example, a human-like memory architecture for LLM-based conversational agents enables autonomous memory recall and dynamic memory consolidation to improve response generation \cite{hou_my_2024}. Additionally, the Long-term Dialogue Agent framework can enable more personalized and contextually relevant responses over extended interactions \cite{li_hello_2024}. A similar advanced memory design can be applied with an agent in the context of language learning. An agent with such capabilities could adapt its interactions based on prior user behavior and progress, and dynamically adjust content delivery according to the learner's needs. In addition, based on the findings in Section \ref{subsubsec:tone_appearance_bodymovement} and criteria from \citet{adjagbodjou_envisioning_2024}, we suggest the value of cultural references and contextual relevance in the adaptive learning experience. For instance, an agent could adjust its dialogue to reflect the cultural nuances of the learner's cultural background, and new environment, offering more meaningful and relatable interactions in the context of language learning.

\subsection{Enhance Agent-Learner Connection}
\label{subsec:design_implications_enhance_agent_learner_connection}
\textit{Design Implication 2: Incorporate user-preferred personas, tone, voice, appearance, and body movement in LLM-based agents to strengthen emotional connections and support language learning.}

Participants in our study exhibited distinct preferences for the agent’s persona, tone of response, voice inflections, appearance, and body movement, which may potentially affect their learning experience with the agent. While previous research has mixed opinions on the impact of an agent's non-verbal features on learning outcomes \cite{petersen_pedagogical_2021, sonlu_effects_2024}, offering customizable options for users may be a beneficial approach. 

\noindent
\textit{Design Implication 3: Integrate advanced systems to effectively use verbal and non-verbal cues to provide emotional support to the learners. }

Emotional support requires sophisticated techniques to accurately interpret learners' intentions and offer contextually appropriate support \cite{dincer_common_2013}. For example, the agent should be able to distinguish between pauses that indicate the learner is thinking and those that suggest the learner is struggling to respond, adapting its behaviour accordingly by offering encouragement or rephrasing questions. Using machine learning techniques to detect user emotions such as silence, hesitation, or frustration could potentially enable the agent to deliver timely and suitable emotional support. For example, a recent study by \cite{siyan_using_2024} has demonstrated the potential of using machine learning and prompt engineering to detect anxiety from students' voices, allowing the agent to offer adaptive, empathetic responses to enhance language learning.

\subsection{Cultivate Long-Term Engagement and Continued Use}
\textit{Design Implication 4: Design LLM-based agents in social VR to facilitate long-term language learning essential for improving fluency. }

Based on the findings in Section~\ref{subsec:continued_use}, participants' willingness to continue using the system regularly suggests that LLM-based agents in VR hold significant potential for supporting long-term language learning. However, this potential is contingent on the system’s ability to adapt to the learner’s progress, sustain engagement, and mitigate issues like conversation flow breakdowns and VR-related usage fatigues. For example, incorporating preferred personality traits or non-verbal cues into the agent can potentially increase engagement \cite{sonlu_effects_2024}. Additionally, gamification is a well-established approach to enhance engagement \cite{koulouris_me_2020, cheng_teaching_2017,hsu_spelland_2023,seow_lingoland_2023}. Most recently, the use of LLM-agents as peers in VR classrooms has been shown to promote engagement and classroom participation \cite{class_meta}.

\noindent
\textit{Design Implication 5: Use Open and Scalable VR Platforms for Sustained Long-Term Development and Studies. } 

Some existing studies have utilized commercially available open and scalable platforms like Second Life \cite{peterson_efl_2012} and Metaverse \cite{cantone_contextualized_2023}. We recommend using similar platforms for sustained development. In our study, we used VRChat, which provides the necessary flexibility for long-term development and broader public usage. Through VRChat's OSC \cite{vrchat_osc_2024} and Unity \cite{unity_unity_2024}, we can develop customizable avatars and virtual worlds, control agent personas, voice tones, facial expressions, and body movements. This scalability ensures that the system can evolve and support long-term studies over time. However, the deployment of AI-agents in social VR platforms like VRChat is a grey area because some view them as modifications that are not supported or permitted according to the platform guidelines\footnote{\url{https://hello.vrchat.com/community-guidelines}}. Thus, the use of AI-agents in social VR requires future considerations of how can we enable more open and scalable platforms for sustainable long-term development and exploration of AI-agents in social VR.

\section{Limitations \& Future Work}
We discuss the limitations of our study and offer suggestions for future work on embodied LLM-based agents for language learning.

\textit{Widening participation}: The participant pool consisted primarily of graduate students with relatively high English proficiency, most of whom were from China. This demographic specificity limits the generalizability of our findings. Future research should investigate how these results apply to a more diverse demographic, including learners with lower language proficiency and varied cultural backgrounds. In addition, the system could be integrated into existing language learning programs offered by educational institutions and immigrant support organizations. This would allow for a more systematic and structured learning experience. Partnerships with these organizations could help in customizing the system further to address the specific needs of this user group.

\textit{Refining LLM Prompt Engineering for Language Learning Tasks} LLM’s performance is very sensitive to prompts. Our current prompt engineering approach is largely based on trial and error due to lack of previous research on prompt engineering for language learning tasks, which may not be the most effective method for optimizing LLM performance. As more prompt frameworks become available \cite{khattab_dspy_2023}, and the capabilities of LLMs continue to evolve, more efficient techniques for optimizing prompts are likely to become available. This could reduce the impact of specific prompt wording, shifting the focus towards refining the reasoning and task breakdown logic for complex, multi-turn conversations. Further research is necessary to develop more systematic workflows for enhancing LLM performance in these contexts.

\textit{Exploring ECA Features for Language Learning in social VR} Given the limited research on how the verbal and non-verbal features of ECAs affect learning outcomes in social VR (as discussed in Section \ref{subsec:design_implications_enhance_agent_learner_connection}, particularly in the context of language learning, further exploration is necessary to better understand and optimize these interactions. Some initial work has been done on the effects of embodiment and personality traits in LLM-educational agents \cite{sonlu_effects_2024} that could be implemented in an embodied LLM-agent for language learning. More research needs to be done to validate the findings in the context of language learning in social VR. 

\textit{Optimizing LLM Selection for Task-Specific Performance} Achieving the design goals of ELLMA-T may require the use of different LLMs tailored to specific tasks. For conversational tasks, GPT-4o \cite{openai_hello_2024} is recognized as the best-performing model, capable of understanding user sentiment and maintaining engaging dialogues \cite{openai_hello_2024}. Other models may be more suitable for summarization and feedback due to their reliability \cite{caines_application_2023}. Preliminary studies suggest that GPT-4o is the best-performing model we can use for both tasks in this study. In future work, we may use or combine different models to leverage the strengths of various models and optimize the system's performance for future improvements.

\textit{Validating LLM Assessment of Language Proficiency}: We recognize that the use of CEFR criteria \cite{coe_cefr_2024} to prompt the LLM in our system is not entirely precise and serves primarily as a rough guide for the LLM to adjust language difficulty for following dialogues. Whether the dialogue content generated by the LLM accurately reflects the intended difficulty level remains to be rigorously validated in future studies.

\textit{Implementing Content Filters for LLM Responses} Additionally, we did not implement a filtering layer to screen the LLM's responses before delivery, which could lead to hallucinations or potentially harmful content. Although our initial tests showed that GPT-4 generally provided correct responses in grammar correction tasks, existing studies indicate that it may still produce inaccurate results or introduce stereotypes \cite{siyan_using_2024}. Future work will include the evaluation of LLM responses by using the LLM Judge framework \cite{zheng_judging_2023} or CoMPosT framework \cite{cheng_compost_2023} to detect stereotypes.

\textit{Evaluating Long-Term Educational Impact} Lastly, this preliminary study involved only 15-minute interaction sessions, and we did not measure the system's effectiveness in terms of actual language learning outcomes. To fully assess the educational impact of ELLMA-T, we plan to conduct longitudinal studies with longer learning sessions. In addition, these studies will provide a more comprehensive evaluation of user engagement over time and help identify key factors that contribute to sustained use and long-term language learning outcomes. 

\section{Conclusion}
With this study, we contribute to the growing body of research on ECAs and language learning by demonstrating the potential of combining situated learning principles with LLM-based agents in social VR environments. In particular, we contribute a system design for an LLM-agent for language learning (ELLMA-T), findings from a qualitative evaluation study (N=12), and design implications based on our findings. Our findings highlight the importance of integrating nuanced non-verbal cues and customizing agent interactions to address the diverse needs of adult language learners. The system also provides benefits such as reducing speaking anxiety and providing potential flexible learning modes. However, our study also highlighted areas for improvement including the need for more sophisticated emotional support, better handling of communication obstacles, and refined conversation flow to maintain long-term user engagement. These findings offer implications for future development, positioning embodied LLM-based agents as a promising tool for personalized and context-aware language learning.

\bibliographystyle{ACM-Reference-Format}
\bibliography{CHI25_main}

\newpage
\onecolumn
\appendix

\section{Appendix}
\subsection{Single System Prompt}
\label{appendix:single_prompt}
The following content is the single prompt:
context = """You are a friendly, and very patient language partner, who also understands the user's cultural background.
         Ensure that your responses do not contain greetings like Hello."""

        steps = """Here are the steps for the role-play language practice:

 1. **Initial Assessment**:
    - Assess the user's spoken language level by conducting an initial free-form conversation.
    - Follow the criteria by the CEFR from this website: [CEFR Criteria](https://www.coe.int/en/web/common-european-framework-reference-languages/table-3-cefr-3.3-common-reference-levels-qualitative-aspects-of-spoken-language-use).
    - Assess only once during the whole conversation.
    - Tell the user the assessment result and adjust your vocabulary difficulty levels based on the user's level.

 2. **Scenario Selection**:
    - Ask the user if they have a specific scenario in mind for practicing.
    - If not, suggest three different types of real-life scenarios and have the user pick one.
    - Each scenario should be different. Ask for a new scenario only when the current role-play is over.
    - Example: In one scenario, you are a colleague with the user and you work at the same company. You invite the user to have lunch together. Discuss what food to eat and about food from your culture.

 3. **Role-Play Execution**:
    - Introduce the user to around 5 new vocabularies for any scenario. If the user doesn't know the meaning of the vocabulary, explain and practice pronunciation with them.
    - If the user's response doesn't make sense, clarify and let the user repeat or rephrase.
    - If the user has difficulty expressing themselves, provide scaffolding by encouraging them, like "You're doing great" or "Come on, don't give up", or offering support.
   - If the user's response is too short, provide scaffolding by suggesting an example reply and asking the user to practice with you.

 4. **Engagement**:    
    - Make the conversation interesting and engaging.
    - Continue the role-play conversation for every topic unless the user explicitly wants to end the topic.
    - After each user response, ask engaging follow-up questions or introduce new related topics to keep the conversation going.

 5. **Feedback**:
    - When the role-play is over, give the user balanced feedback, taking into account the difficulty level of the conversation.
    - Summarize the vocabulary, grammar, or sentences they have learned for future revisit.

 6. **Instructions to Continue**:
    - If the user responds, acknowledge it and ask a follow-up question or provide an engaging prompt to continue the conversation.
    - Only end the role-play if the user explicitly states they want to stop or change the topic.

\subsection{Multi-task Prompt}
\label{appendix:multi-task sample promt}
\begin{lstlisting}
1. Introduction
{"role": "system", "content": "You are a friendly, patient, and light-hearted English language tutor."},
{"role": "user", "content": " Greet me and chat with me to get to know me better, such as my name, cultural background, and why learning English, ask one question at a time and wait for my response to continue. If you know my cultural background, greet me in the language of that culture. "}
2. Assessment
 {"role": "system", "content": "You are an English language tutor assessing the user's language proficiency. You are familiar with the criteria of the CEFR from this website: [CEFR Criteria](https://www.coe.int/en/web/common-european-framework-reference-languages/table-3-cefr-3.3-common-reference-levels-qualitative-aspects-of-spoken-language-use)."},
 {"role": "user", "content": "Wrap up our previous conversations in one sentence, here are the previous conversations: {user_info_conversation}. Transit to assessing my language level. Ask me to describe a memorable experience and tell me the assessment result based on the criteria by the CEFR from this website: [CEFR Criteria](https://www.coe.int/en/web/common-european-framework-reference-languages/table-3-cefr-3.3-common-reference-levels-qualitative-aspects-of-spoken-language-use)."}
3. Generate Role-Play Scenarios
{"role": "system", "content": "You are a friendly, patient, and light-hearted English language tutor. You are asking the user to choose a role-play scenario."},
{"role": "assistant", "content": "Suggest me three real-life scenarios I can practice. For example, ordering food at a restaurant, a job interview, or traveling in an English-speaking country. Do not list them in numbers."}
4. Role-Play Conversation Practice
{"role": "system", "content": f"""You are a friendly, patient, and light-hearted English language tutor engaging in a role-play scenario: {scenario} to help the user practice listening and speaking skills. 
Adjust your language difficulty level by the assessment result from previous conversations: {assessment}. Think how you can help a learner to learn better."""},
{"role": "assistant", "content": f"""Explain to the user that you're helping them practice speaking English by conducting role-play conversations. Start the role play. You're at a {scenario}. Start by describing the scene and playing the role. Let the user continue the conversation.
If the user doesn't know the meaning of a vocabulary, explain and practice pronunciation with them.
If the user's response doesn't make sense, clarify and let the user repeat or rephrase.
If the user has difficulty expressing themselves, provide scaffolding by encouraging them, like "You're doing great", "Come on, don't give up", or offering support.
If the user's response is too short, provide scaffolding or suggest an example reply and ask the user to practice with you."""}
5 Providing Feedback
{"role": "system", "content": "You are a friendly, patient and light-hearted English language tutor. You are providing feedback on the user's English conversation practices."},
{"role": "assistant", "content": """Based on previous role-play conversation practices: {role_play_conversations}. Summarize the vocabulary, grammar, or sentences they have learned for future revisit.
Feedback should be in the following format:
**GENERAL FEEDBACK**: Assess performance given the lesson, name one thing the student did really well, and one thing the student could improve on.
**ADVICE MOVING FORWARD**: Give students advice on how to apply the lesson in the real world."""}
\end{lstlisting}

\section{Interview Questions}
\label{appendix:interview_questions}

The following questions were asked during the semi-structured interviews conducted as part of the user study:

\begin{table}[h]
\centering
\begin{tabular}{|l|p{10cm}|}
\hline
\textbf{Focus Area} & \textbf{Interview Question} \\ \hline
Initial Thoughts    & Tell me your initial thoughts about the interaction with ELLMA-T. \\ \hline
Skill Assessment    & How well did it assess your skills? \\ \hline
Feedback Quality    & What did you think of the feedback? What was helpful or not? \\ \hline
Comfort Level       & How comfortable do you feel interacting with the agent? \\ \hline
Engagement          & How engaging do you find the interactions with the agent? \\ \hline
Interaction Mode  & Would you prefer interacting with the agent alone or in a group setting within VRChat? Why? \\ \hline
Likelihood of Continued Use & How likely are you to continue using this system over a long period? \\ \hline
\end{tabular}
\caption{List of interview questions used in the study.}
\label{tab:interview_questions_appendix}
\end{table}

\end{document}